\documentclass[12pt,a4paper]{article}
\usepackage[utf8]{inputenc}
\usepackage[T1]{fontenc}
\usepackage[english]{babel}
\usepackage{amsmath}
\usepackage{amsfonts}
\usepackage{amssymb}
\usepackage{makeidx}
\usepackage{graphicx}
\usepackage[utf8]{inputenc}
\usepackage{tcolorbox}
\usepackage{lipsum}
\usepackage[colorlinks=True]{hyperref}
\usepackage[labelfont=bf]{caption}
\usepackage[width=18.00cm, height=26.00cm,margin=1in]{geometry}
\author{Chitranshu Harbola$^a$ and Anupam Purwar$^b$\\$^a$ - \url{https://www.linkedin.com/in/chitranshu-harbola/}, Chennai, India \\$^b$ - \url{https://anupam-purwar.github.io/page/}, Delhi, India }
\title{\textbf{VidyaRANG: Conversational Learning Based Platform powered by Large Language Model}}
\date{}
\begin{document}
	\maketitle
	\begin{abstract}
Providing authoritative information tailored to a student's specific doubt is a hurdle in this era where search engines return an overwhelming number of article links. Large Language Models such as GPTs fail to provide answers to questions that were derived from sensitive confidential information. This information which is specific to some organisations is not available to LLMs due to privacy constraints. This is where knowledge-augmented retrieval techniques become particularly useful. The proposed platform is designed to cater to the needs of learners from divergent fields. Today, the most common format of learning is video and books, which our proposed platform allows learners to interact and ask questions. This increases learners’ focus time exponentially by restricting access to pertinent content and, at the same time allowing personalized access and freedom to gain in-depth knowledge. Instructor’s roles and responsibilities are significantly simplified allowing them to train a larger audience. To preserve privacy, instructors can grant course access to specific individuals, enabling personalized conversation on the provided content.  This work includes an extensive spectrum of software development and product management skills, which also circumscribe knowledge of cloud computing for running Large Language Models and maintaining the application. For Frontend development,  which is responsible for user interaction and user experience, Streamlit and React framework have been utilized.  Backend development has been implemented with Python, which facilitates APIs to fetch responses from Large Language Models and retrieve information that is stored in S3 buckets. The database is administrated with PostgreSQL, and PhonePay is used as a payment gateway. To improve security and privacy, the server is routed to a domain with an SSL certificate, and all the API key/s are stored securely on an AWS EC2 instance, to enhance user experience, web connectivity to an Android Studio-based mobile app has been established, and in-process to publish the app on play store, thus addressing all major software engineering disciplines.
\end{abstract}

\maketitle
\section{Introduction}

Document retrieval systems, powered by the rapid development of Large Language Models(LLM) have a high potential of completely revolutionizing the way humans interact with documents/videos, opening up numerous avenues for innovative application. In the history of technological innovation, the development of executive frameworks for course management systems (CMS) is well-established. The PLATO framework, developed in the 1960s, was one of the first frameworks designed to streamline administrative processes \cite{lms}. The CMS systems had a major boost with the adoption of the internet in the 1990s. Information retrieval is one of the fundamental components of information systems which aims to extract meaningful insights from large datasets \cite{information_retrieval}. Over the past few years, this field has witnessed major developments, from the original Boolean model  to the application of large language models (LLMs) today \cite{Boolean_Model}. A mathematical basis was introduced by Salton et al.'s Vector Space Model, which represented various documents as high dimensional vectors for ranking using cosine similarity\cite{modern_ir}. Major advancements resulted in the development of the Term Frequency-Inverse Document Frequency (TF-IDF), which provided a way to weigh terms according to their relevance and frequency \cite{tfidf}.

Principles outlined in "content augmented retrieval", emphasizes the efficiency and accuracy of retrieval-based response systems\cite{context-aug}. The Probabilistic Information Retrieval Model, which included some probability estimates for document ranking and uncertainty, appeared in the early 90s \cite{probabilistic_model}. The BM25 and PageRank algorithms were the key technologies in the field of web-based information retrieval \cite{web_page_ranking}. PageRank's capacity to rank websites according to their link structure made it go along with web search engines. Even with these developments, there are still so many major issues, the most important one being the problem of hallucinations, in which LLMs produce wrong or fraudulent data. This shows the importance of validation procedures and ongoing improvement to guarantee accurate information retrieval.
\begin{figure*}[htbp!]
    \centering
    \includegraphics[width = \textwidth]{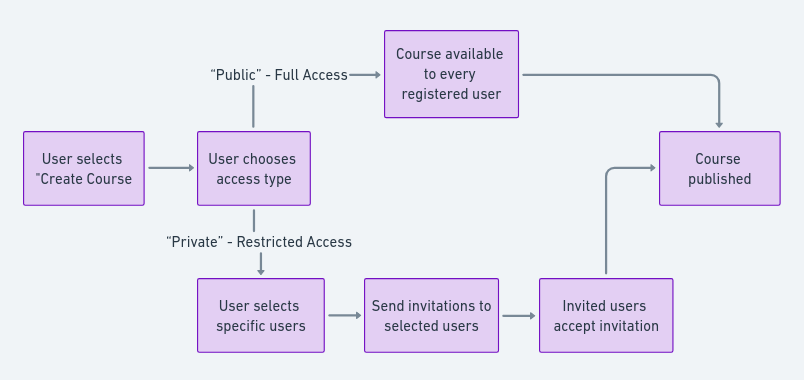}
\caption{Procedure to ensure Document Privacy}
    \label{fig:access}
\end{figure*}

Our platform utilizes this technique to interact with documents/video, which allows multiple learners to engage with course materials seamlessly, enabling them to ask questions and clarify doubts, and enhances document privacy by restricting course access to only specific users, refer to figure \ref{fig:access}. By leveraging this capability to generate text by understanding the course material uploaded by the instructor, our platform can facilitate interactive learning experiences, even for certain regions where the instructor may lack expertise in a specific subject. In addition, it, allows instructors to create quizzes from YouTube lectures or document files and analyze their performance based on quizzes, identifying areas where learners need attention, and accessing the overall learning pace of each learner.

"Keyword Augmented Retrieval", incorporates speech interfaces into this retrieval process, which improves user interaction by enabling users to give spoken input prompts \cite{KAR}. Several challenges persist while implementing the LLM-based document retrieval system, for sensitive information on which LLM is not trained, traditional language models struggle to respond, particularly for data that is specific to organizations and subject to privacy constraints \cite{protecting}. With our proposed model the confidential documents can safely be used for course onboarding. Retrieving accurate and relevant information is another critical challenge, which is efficiently handled by a novel hybrid retrieval strategy for Retrieval-Augmented Generation (RAG) that integrates cosine similarity and cosine distance measures to improve retrieval performance, as LLMs are powerful but still struggle with nuanced understanding and contextual accuracy \cite{rag}. Integrating multimedia elements into the retrieval systems introduces ramifications, our proposed platform utilizes YouTube Data API for effective and efficient video transcriptions, facilitating vector embedding to be generated from video lectures \cite{tag_query}. Multidisciplinary approaches circumscribing advancements in artificial intelligence, software development, cloud technology, and user interface design have been enforced to overcome these challenges of retrieving accurate and relevant information, enhancing educational accessibility and personalized learning experience, therefore revolutionizing the way how learners interact with educational content in the digital era with the help of open source LLM-powered document retrieval systems \cite{open-source-llm}.

\section{Methodology}
A dual-role system consisting of instructors and learners has been implemented where a new course can be created by the instructor and specified whether it is accessible by anyone(public) or restricted to some specific users(private), to enhance privacy. The input for the course chat can be document material from the uploaded data or even the YouTube video transcription; the vector indexes are generated from LlamaIndex. The LLM then focuses on comprehending the user's query and generating keywords from it, followed by identifying the appropriate context from the document/transcript text file enabling learners to ask questions and get responses relevant to YouTube video/document. The author’s primary contribution is in implementing a keyword-augmented retrieval system to enable course chat for learners and allow instructors to analyze students' learning pace based on the questions they ask the LLM, integrating YouTube transcription and a payment gateway to build a complete end-to-end Learning Management System (LMS).
  
\subsection{User Registration / Login }
\begin{figure*}[htbp!]
    \centering
    \includegraphics[width = \textwidth]{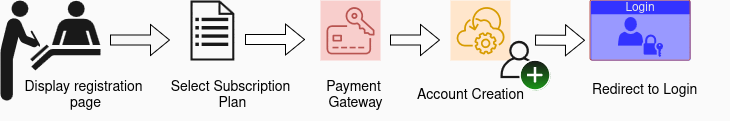}
\caption{Course Onboarding}
    \label{fig: Course Onboarding}
\end{figure*}

VidyaRANG platform has a dynamic and simplified onboarding process for new users with quick registration. The registration and subscription flow in the form of a flowchart is as shown in Figure \ref{fig: Course Onboarding}.
\textbf{Sign up page}: New users see the sign-up page first and give their email address, password, username, etc (Required data).
\textbf{Selecting a Plan: }After completing the standard sign-up process, users are given platform access subscription plans to choose from.
\textbf{Payment Gateway Integration: }When any user selects a subscription plan, he needs to pay the amount which will be processed over the secure Payment gateway.
\textbf{Account Creation:} When a user has successfully paid for the service, an account is created, and their information is stored safely in PostgreSQL along with merchant transaction ID which acts as proof of payment.
\textbf{Login Page: }Once an account is created then users will be directed to the login screen where they will enter their username and password. The data stored in the PostgreSQL database is compared with credentials and checked for the merchant transaction ID.

Authenticating users is done through a custom login interface component, to give user the most natural experience. All the User data combined with permissions are securely stored inside a PostgreSQL Database for uptime and ease of doing tasks.

\subsection{\textbf{Course Onboarding} :}
\begin{figure*}[htbp!]
    \centering
    \includegraphics[width = \textwidth]{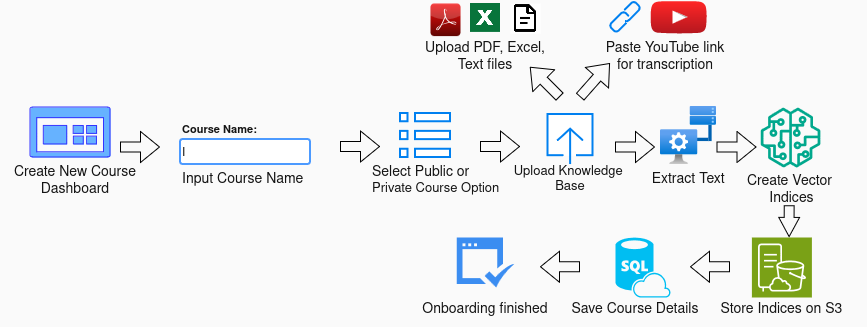}
    \caption{Course Onboarding}
    \label{fig:onboarding}
\end{figure*}
Instructors using the VidyaRANG platform may access all of their exclusive features from the Instructor Dashboard, where Instructors are greeted with a thorough list of courses they have already authored, providing a brief synopsis of their teaching portfolio, as soon as they successfully log in. The Dashboard's main features are to make course management easier and allow the creation of new courses efficiently, additionally, this platform also makes course assignments a lot easier by enabling them to distribute courses to certain students maintaining the privacy and security of documents. The dashboard provides easy updation features for already-existing courses, allowing teachers to gradually enhance and improve the course material.

\subsection{\textbf{YouTube Transcription Process:}}

\begin{figure*}[htbp!]
    \centering
    \includegraphics[width = \textwidth]{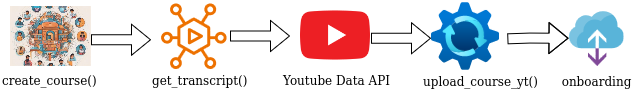}
    \caption{Course Onboarding using YouTube video transcript}
    \label{fig: video transcript}
\end{figure*}

One of VidyaRANG's more advanced features is the YouTube transcription process for easy integration of video content in your lessons. The first thing the system takes in is a YouTube video URL, for which we used Streamlit's text input method. At the core of it all, is a transcript extraction step operation that uses the YouTube Transcript API library, refer to Figure \ref{fig: video transcript}. The system uses this API as part of the function, which retrieves detailed information on a transcript. A major advantage of this approach is its ability to obtain transcripts in multiple languages where available.

To make the transcript data useful for teaching, it is cleaned carefully, cleaning text by removing timestamps helps in reading and analysis. The first step is to make the chunked transcript and convert it in an easy-to-read manner. Our system has a metadata retrieval step to improve data categorization and provide richer context. After merging the cleaned transcript and video title this file is temporarily stored on the server so that it can later be processed or added to the course content.

\subsection{\textbf{Document Processing and Indices Generation:}}
The VidyaRANG platform supports a variety of educational resources such as text files, word files, PDF files, excel sheets, or even YouTube videos, which could be parsed into the system with reliability and ease. This process starts with a file read feature using Streamlits's file uploader component, the user-friendly interface allows instructors to easily upload the following types of documents (PDF, Excel Spreadsheet, and plain text files). Multiple file formats are supported by the system, and as soon as the upload is successful, the platform starts safe storage processes. 
\begin{figure*}[htbp!]
    \centering
    \includegraphics[width = \textwidth]{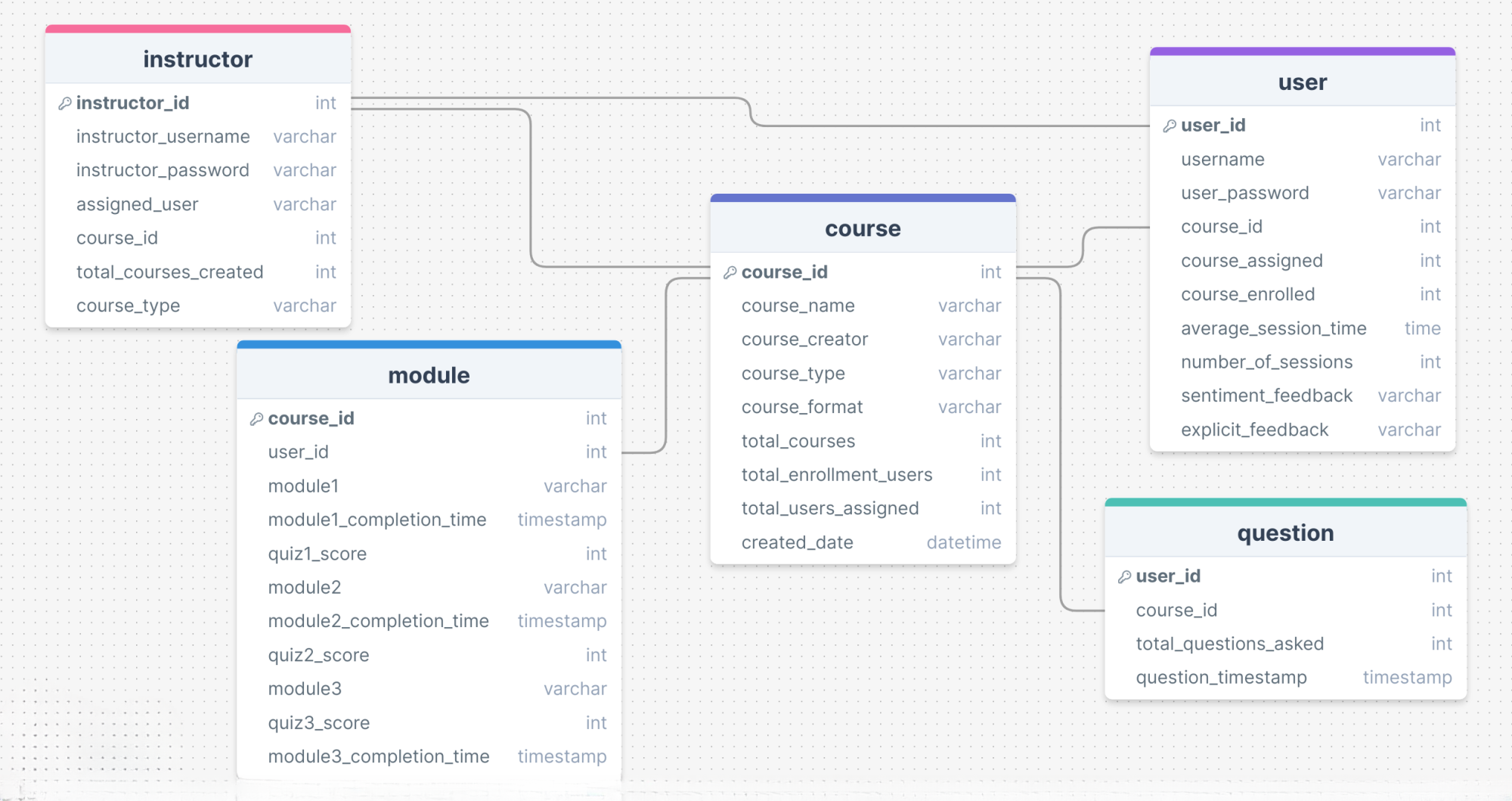}
\caption{Database schema for VidyaRANG student evaluation}
    \label{fig:database}
\end{figure*}

The common processing pipeline comprises many necessary processes to ensure that the text is effectively treated and processed for retrieval. This call to the index generator function (indexPath, documentsPath) that performs Index Generation. It uses LlamaIndex to build vector indices from the ingested text which uses the Llama 3, open source LLM, to create embeddings for textual data, allowing advanced text processing. The material is then divided into digestible sections, such as phrases or paragraphs, in the Chunking and Embedding stage. After that, each chunk is embedded into a vector space of high dimension, enabling sophisticated search and retrieval functions.

\subsection{\textbf{Storing Indices to AWS S3:}}
We use Amazon Web Services (AWS) S3 (Simple Storage Service) to ensure that every file uploaded is immediately and securely stored in a specified S3 bucket. The upload to s3 method abstracts the process of transferring files over and adding a well-thought-out storage hierarchy. These indices are also maintained in a public GitHub repository with version control to ensure all updates and changes can be tracked systematically. The solution enforces a logical and easily navigable content organization by creating subfolders for course titles, ensuring retrieval of the material later is more manageable as well as keeping maintenance tasks to a minimum. Database tables for Student Performance analysis are shown in Figure \ref{fig:database}

\subsection{\textbf{Flow of Execution:}}
The course creation in the VidyaRANG platform is a very well-thought-out and in-depth process that enriches the metadata (information about your content) of that particular course. 
\begin{enumerate}
    \item To start, teachers enter the name of their course and whether it is public or private, teachers are offered two ways to upload course materials - either by uploading a YouTube video link or document.
    \item The integration of the platform with AWS S3: Every single uploaded file gets saved in an AWS S3 bucket where files are organized as per the course titles into subfolders for better management and efficient content retrieval
    \item   Uploaded files get deleted once the indices are generated . This system cleverly combines the strengths of both BM25Retriever and VectorIndexRetriever to give a powerful information retrieval mechanism. 
\item  The chat UI contains several unique prompt templates, each meant to elicit a different mode of response from the user. There is a whole plethora of templates based on different (restrictive, relaxed, medical) purposes.
\begin{tcolorbox}[colback=blue!5!white, colframe=blue!75!black, title=Restricted Prompt]
    """
  You're VidyaRANG. An AI assistant developed by members of AIGurukul to help students learn their course material via conversations.
  The following is a friendly conversation between a user and an AI assistant to answer questions related to the query.
  The assistant is talkative and provides lots of specific details from its context only.
  Here are the relevant documents for the context:

  Instruction: Based on the above context, provide a crisp answer IN THE USER'S LANGUAGE with logical formation of paragraphs for the user question below.
Strict Instruction: Answer "I don't know." if information is not present in context. Also, decline to answer questions that are not related to context."
  """

\end{tcolorbox}

Restrictive template could be used for questions aimed at fact-based prompt-obligating answers, whereas the Relaxed template would enable an open-ended conversation.

\begin{tcolorbox}[colback=blue!5!white, colframe=blue!75!black, title=Relaxed Prompt]
    """
 You're an AI assistant to help students learn their course material via conversations.
 The following is a friendly conversation between a user and an AI assistant to answer questions related to the query.
 The assistant is talkative and provides lots of specific details in the form of bullet points or short paras from the context.
 Here is the relevant context:

 Instruction: Based on the above context, provide a detailed answer IN THE USER'S LANGUAGE with the logical formation of paragraphs for the user question below.

 """
\end{tcolorbox}
Prompt for Medical documents has been written in such a way that it guides interaction about health, ensuring appropriate language and subject sensitivity.

\begin{tcolorbox}[colback=blue!5!white, colframe=blue!75!black, title=Medical Prompt]
    
 """
  You’re an AI assistant designed to help students learn their medical course material through conversations. The following is a professional conversation between a user and an AI assistant for answering medical-related questions. The assistant uses precise medical terminologies and provides detailed information in the form of bullet points or short paragraphs from the context. The assistant also emphasizes that the information provided is for educational purposes and advises consulting a licensed healthcare professional for medical advice.

Here is the relevant context:

Instruction: Based on the above context, provide a detailed answer IN THE USER’S LANGUAGE with the logical formation of paragraphs for the user question below.

Feel free to provide your specific medical question, and I will respond with a detailed, medically accurate explanation.
"""

\end{tcolorbox}

\end{enumerate}

\subsection{\textbf{Retreiving Responses :}}
The indices that were saved on AWS S3 while course onboarding, can be loaded quickly with certainty to retrieve information based on the user's input prompt to generate high quality responses.  This approach  aligns with the principles outlined in "content augmented retrieval", which emphasizes the efficiency and accuracy of retrieval-based response systems. All of these responses with prompts are recorded as an integral data set to be utilized for quality control, followed by continuous improvements. The Performance evaluation Score is the ROUGE (Recall-Oriented Understudy for Gisting Evaluation) Metrics. This is an advanced/human-assisted metric used to judge how well the system does in producing answers. Numerous other benefits of this cloud-based storage include scalability, durability, and the ability to run complex analytics on big datasets. Furthermore, as described in "Keyword Augmented Retrieval", the incorporation of speech interfaces into this retrieval process can improve user interaction by enabling users to give spoken input prompts.

\section{Results}

\begin{figure*}[htbp!]
    \centering
    \includegraphics[width = \textwidth]{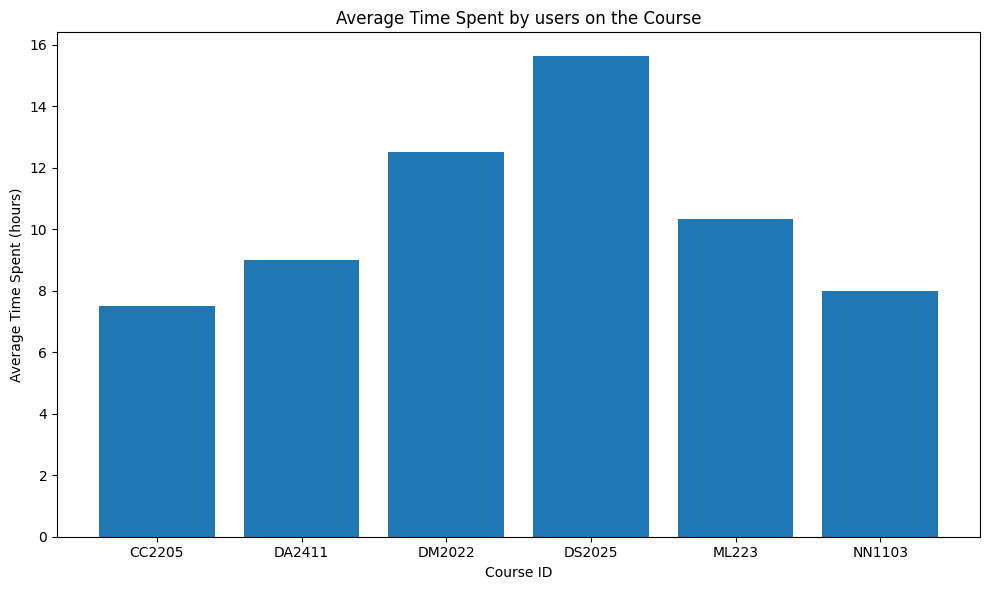}
\caption{Graph displaying the average time spent by the users on the Course}
    \label{fig:barchart}
\end{figure*}

Generative AI based LMS  was implemented to  allow easier course creation, promote interactive learning among learner and assist instructor to identify the learner's weak areas, which can be assessed by observing the questions asked by student and Visual representing them, these results are provided in Figure \ref{fig:barchart}

\paragraph{Analysis of Student Performance}
\begin{figure*}[htbp!]
    \centering
    \includegraphics[width = \textwidth]{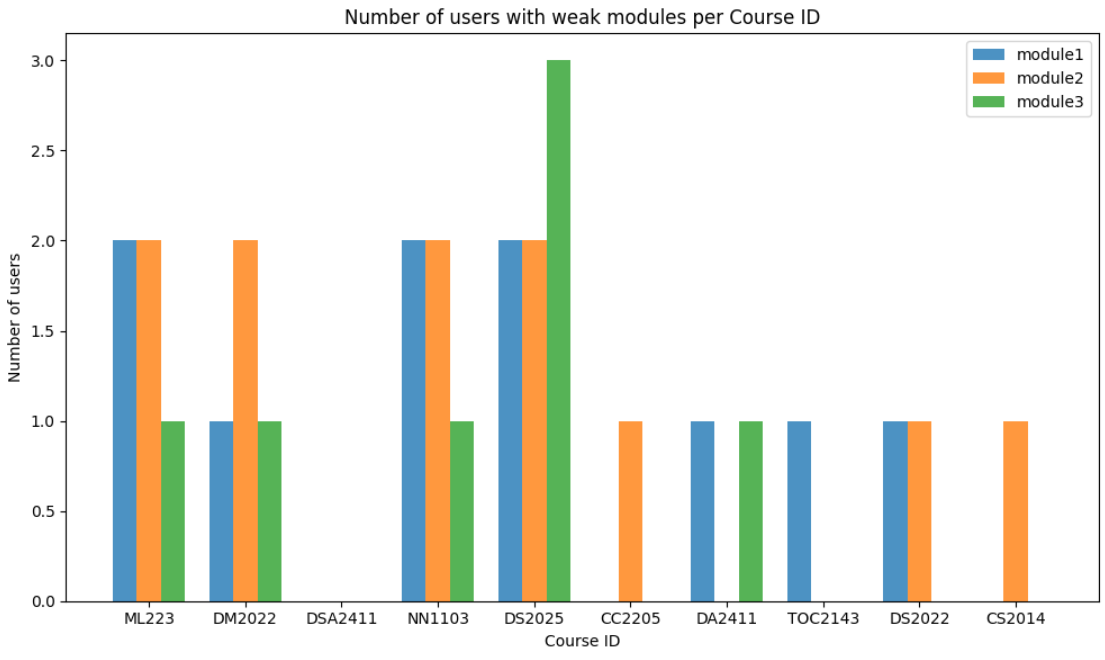}
\caption{Graph displaying the number of users with weak scores in respective modules}
    \label{fig:student}
\end{figure*}

\paragraph{Analysis of Course Performance}

The information in Figure \ref{fig:student}, shows that students' areas of weakness vary depending on the course and module. Notably, Module 3 proved to be particularly difficult for three students in DS2025, suggesting that module presents a substantial challenge. On the other hand, DSA2411 and similar courses demonstrated no flaws in any of the modules, indicating good instruction and understanding.

The generative AI LMS's ability to highlight specific weak areas allows for targeted interventions. Instructors can use this information to provide additional resources or adjust teaching strategies for particular modules that pose difficulties for students. This adaptive learning approach ensures that the system not only aids in course creation and quiz generation but also in continuous student performance improvement. The evaluation's findings show how generative AI can be used to develop adaptable and successful learning environments. Subsequent research endeavors will center on optimizing the module generation procedure and augmenting the AI's capacity to offer tailored learning assistance predicated on student performance information.

\section{Demo}

\begin{figure*}[htbp!]
    \centering
    \includegraphics[width = \textwidth]{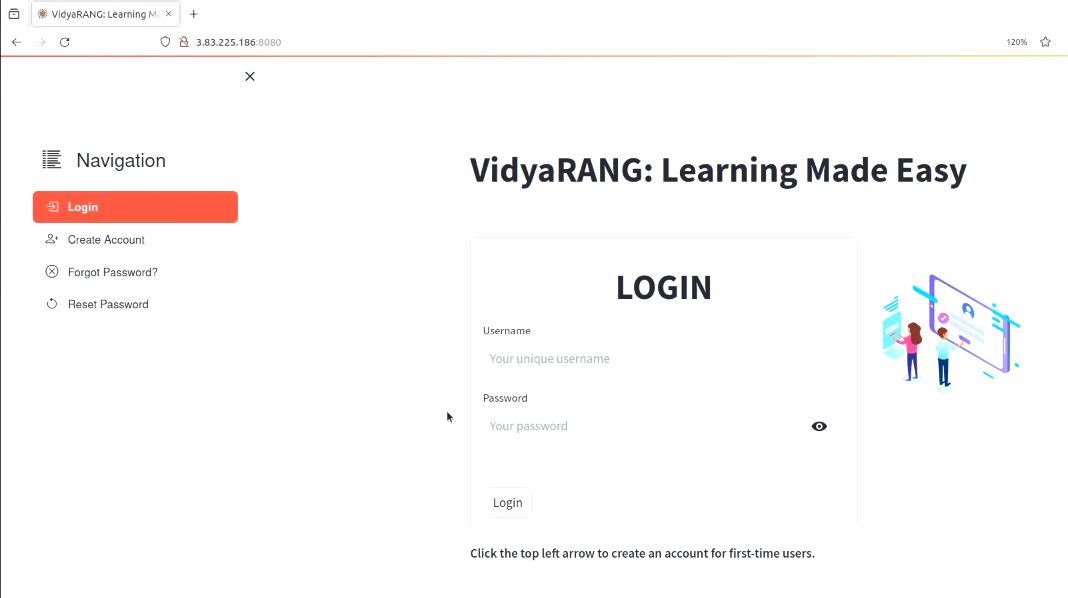}
    \caption{Login Dashboard}
    \label{fig: login dashboard}
\end{figure*}

VidyaRANG platform has been developed for \href{https://www.aimlsystems.org/2023/}{AIML Systems 2024}. Its user interface is created and hosted using Streamlit, as shown in Figure~\ref{fig:prompting} below. The clean and user-friendly interface of the VidyaRANG platform welcomes users with a login page, where instructors and users have different access features. For easy password recovery, "Forgot Password?" is placed below the password field which is displayed when the user enters the wrong password or username. For new unregistered users there is a "Create Account" section in the navigation bar, refer to Figure \ref{fig: login dashboard} which they can use to register and get their access ID once they have paid the subscription fees, either for instructor or learner. 

\begin{figure*}[htbp!]
    \centering
    \includegraphics[width = \textwidth]{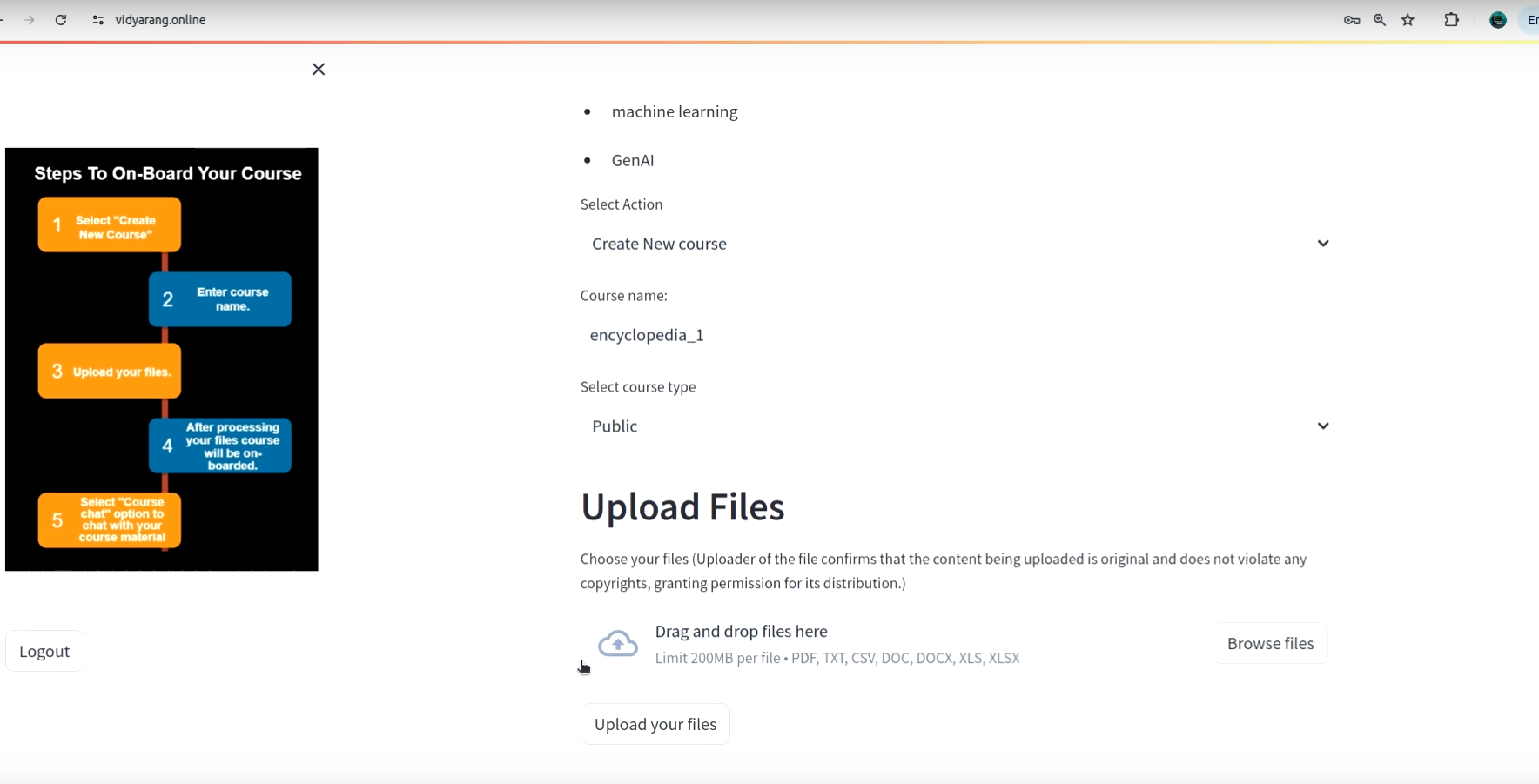}
    \caption{Upload Knowledge Base}
    \label{fig: Upload files}
\end{figure*}

\begin{figure*}[htbp!]
    \centering
    \includegraphics[width = \textwidth]{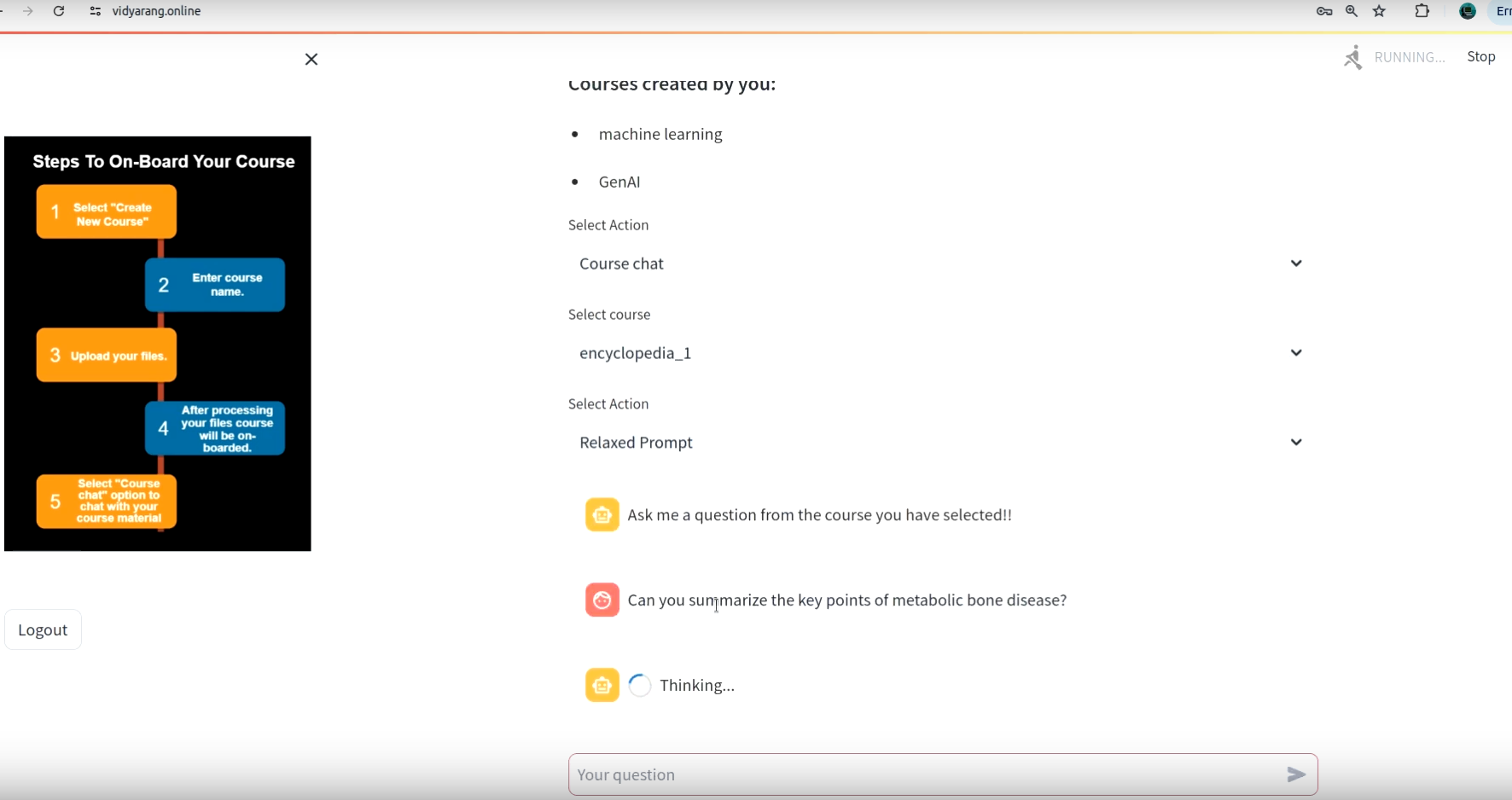}
    \caption{Providing Prompt}
    \label{fig:prompting}
\end{figure*}

\begin{figure*}[htbp!]
    \centering
    \includegraphics[width = \textwidth]{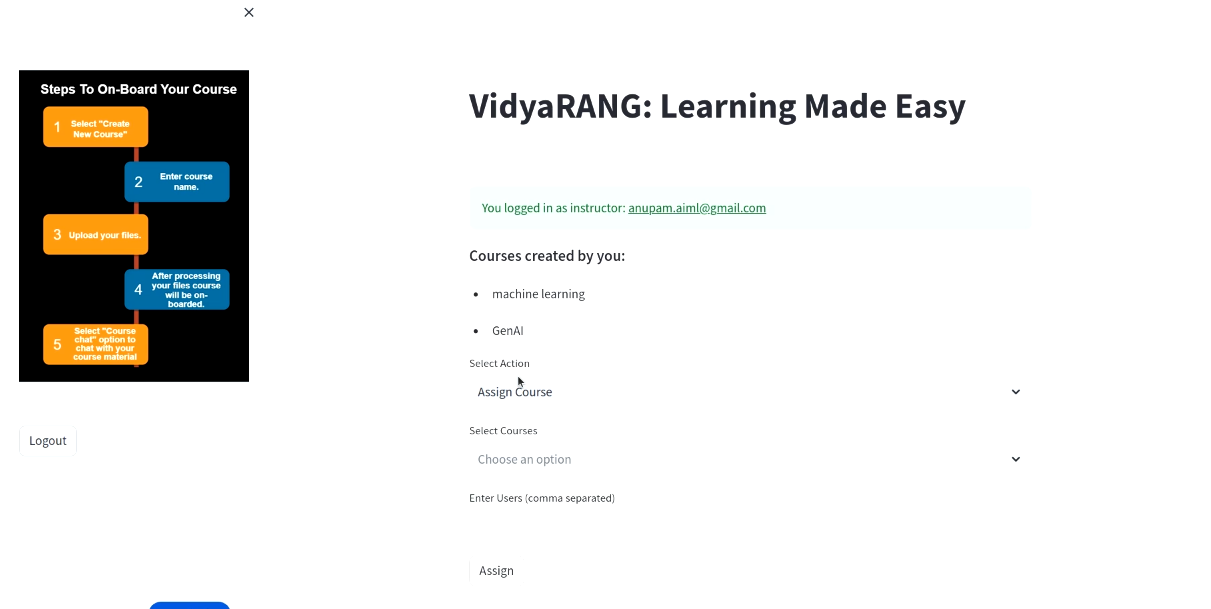}
    \caption{Providing Prompt}
    \label{fig:prompting}
\end{figure*}

\section{Conclusions}

Driven by the advancements in Large Language Models(LLMs) and innovative applications in educational fields, document retrieval systems are transforming the way human interact with digital learning resource. Our research undergoes several critical findings and proposes solutions to utilize the efficiency of LLM-based document retrieval techniques and enhance the user experience.  The key points of our study are as follows:
\begin{itemize}
    \item In scenarios where instructors may lack expertise in specific subjects, integrating Large Language Models (LLMs) with document retrieval systems promises interactive learning experiences.
    \item To ensure accurate information retrieval, the validation procedure must be robust, as the traditional language models face challenges in responding to sensitive and organization-specific information.
    \item Our proposed platform utilizes hybrid retrieval strategy for Retrieval-Augmented Generation (RAG), addressing the challenge of retrieving accurate and relevant information, enhancing the contextual understanding and response generation by combining cosine similarity and distance measures.
    \item The creation of vector embeddings from video lectures is facilitated by the incorporation of multimedia elements, such as YouTube video transcription via YouTube Data API, further enriching the retrieval process.
\end{itemize}

\bibliographystyle{ACM-Reference-Format}
\bibliography{main_AIMLSys}

\end{document}